\documentclass[aps,pre,epsf,superscriptaddress,amsmath,amssymb,amsfonts,twocolumn,showpacs]{revtex4-1}

\usepackage{graphicx}
\usepackage{epsfig}
\usepackage{dcolumn}
\usepackage{bm}
\usepackage{braket}
\usepackage{amsmath}
\usepackage{color}

\begin{document}
\title{Many-Body Expansion Dynamics of a Bose-Fermi Mixture\\ Confined in an Optical Lattice}

\author{P. Siegl}
\affiliation{Zentrum f\"{u}r Optische Quantentechnologien,
Universit\"{a}t Hamburg, Luruper Chaussee 149, 22761 Hamburg,
Germany}  
\author{S. I. Mistakidis}
\affiliation{Zentrum f\"{u}r Optische Quantentechnologien,
Universit\"{a}t Hamburg, Luruper Chaussee 149, 22761 Hamburg,
Germany}
\author{P. Schmelcher}
\affiliation{Zentrum f\"{u}r Optische Quantentechnologien,
Universit\"{a}t Hamburg, Luruper Chaussee 149, 22761 Hamburg,
Germany}\affiliation{The Hamburg Centre for Ultrafast Imaging,
Universit\"{a}t Hamburg, Luruper Chaussee 149, 22761 Hamburg,
Germany}

\date{\today}

\begin{abstract} 

We unravel the correlated non-equilibrium dynamics of a mass balanced Bose-Fermi mixture in a one-dimensional 
optical lattice upon quenching an imposed harmonic trap from strong to weak confinement. 
Regarding the system's ground state, the competition between the inter and intraspecies interaction strength gives rise 
to the immiscible and miscible phases characterized by negligible and complete overlap of the 
constituting atomic clouds respectively. 
The resulting dynamical response depends strongly on the initial phase and consists of an expansion 
of each cloud and an interwell tunneling dynamics.  
For varying quench amplitude and referring to a fixed phase a multitude of response regimes is unveiled, 
being richer within the immiscible phase, which are described by distinct expansion strengths 
and tunneling channels.  

\end{abstract}

\maketitle

\section{Introduction}

Recent experimental advances in ultracold atomic gases offer the opportunity to realize mixtures 
of bosons and fermions with the aid of sympathetic cooling \cite{Inouye,Fukuhara,Hadzibabic,Truscott,Heinze}.   
These mixtures serve as prototypical examples in which the interacting particles obey different statistics \cite{Pethick_book,Lewenstein_book}.  
For instance and in sharp contrast to bosons, $s$-wave interactions among spin-polarized fermions are 
prevented due to the Pauli exclusion principle. 
The complex interplay of Bose-Bose and Bose-Fermi interactions led to numerous theoretical studies in Bose-Fermi (BF) mixtures 
such as their phase separation process \cite{Das,Viverit1}, stability conditions \cite{Cazalilla,Miyakawa} 
and collective excitations \cite{Pu,Liu}. 

Moreover, BF mixtures confined in optical lattices unveiled a variety of intriguing quantum phases 
including, among others, exotic Mott-insulator and superfluid phases \cite{Cramer,Lewenstein1,Zujev,Dutta}, 
charge-density waves \cite{Mathey1,Lewenstein1}, supersolid phases \cite{Buchler,Titvinidze} 
and polaron-like quasiparticles \cite{Mathey1,Privitera}. 
A commonly used model to describe the properties of such mixtures, e.g. pairing of fermions 
with bosons or bosonic holes for attractive and repulsive interspecies interactions, 
respectively \cite{Lewenstein1,Kuklov}, is the lowest-band BF Hubbard model \cite{Albus,Sanpera}. 
A celebrated problem that has been intensively studied concerns 
the effect of the fermions on the mobility of the bosons.  
Heavier or lighter fermions, mediate long-range interactions between the bosons 
or act as impurities, inducing a shift of the bosonic superfluid to Mott transition \cite{Best} 
caused by the contribution of energetically higher than the lowest-band states. 
This behavior indicated that more involved approximations than the lowest-band BF Hubbard model 
need to be considered for the adequate explanation of the superfluid to 
Mott transition \cite{Mering,Luhmann1,Tewari}. 

Despite the importance of the system's static properties, a particularly interesting 
but largely unexplored research direction in BF mixtures, is to investigate their non-equilibrium 
quantum dynamics by employing a quantum quench \cite{Polkovnikov,Andrei}. 
Referring to lattice systems the simplest scenario to explore is the expansion dynamics of the trapped atomic cloud after 
quenching the frequency of an imposed harmonic oscillator.  
Such studies have already been performed mainly for bosonic ensembles unravelling the dependence of the expansion on the 
interatomic interactions. 
For instance, it has been shown that the expansion is enhanced for non-interacting or hard-core bosons \cite{scattering}, 
while for low filling systems a global breathing mode is induced \cite{breath}.  
Detailing the dynamics on the microscopic level a resonant dynamical response has been revealed 
which is related to avoided crossings in the many-body (MB) eigenspectrum \cite{MLY}. 
A peculiar phenomenon, called quasi-condensation, arises during the expansion of hard-core bosons enforcing 
a temperature-dependent long-range order in the system \cite{quasi,quasi2,emergentH,emergentHT,emergentH2}. 
Moreover, the expansion velocities of fermionic and bosonic Mott insulators have been found to be the same 
irrespectively of the interaction strength \cite{MottBF}. 
However, a systematic study of the expansion dynamics in particle imbalanced BF mixtures still lacks. 
In such a scenario, it would be particularly interesting to examine how interspecies correlations, which reflect the initial 
phase of the system \cite{Lai,Imambekov,Imambekov1,Fang,Dehkharghani}, modify the expansion dynamics of the mixture. 
Another intriguing prospect is to investigate, when residing within a specific phase, whether different response regimes can 
be triggered upon varying the quench amplitude. 
To address these intriguing questions hereby we employ the Multi-Layer Multi-Configurational Time-Dependent Hatree Method  
for Atomic Mixtures (ML-MCTDHX) \cite{MLX,Cao_ML} being a multiorbital treatment which enables us to 
capture the important inter and intraspecies correlation effects.   

We investigate a BF mixture confined in an one-dimensional optical lattice with an imposed harmonic trap. 
Operating within the weak interaction regime, we show that the interplay 
of the intra and interspecies interactions leads to different ground state phases 
regarding the degree of miscibility in the mixture, namely to the miscible and the immiscible phases where the bosonic 
and the fermionic single-particle densities are completely and zero overlapping respectively.   
To trigger the dynamics the BF mixture is initialized within a certain phase and a quench from strong 
to weak confinement is performed.  
Each individual phase exhibits a characteristic response composed by an overall expansion of both atomic clouds and 
an interwell tunneling dynamics.  
Referring to the immiscible phase a resonant-like response of both components occurs at moderate quench amplitudes being 
reminiscent of the single-component case \cite{MLY}.  
A variety of distinct response regimes is realized for decreasing confinement strength.  
Bosons perform a breathing dynamics or solely expand while  
fermions tunnel between the outer wells, located at the edges of the bosonic cloud, or exhibit a delocalized 
behavior over the entire lattice.  
To gain further insight into the MB expansion dynamics the contribution of the higher-lying orbitals is analyzed 
and their crucial role in the course of the evolution is showcased. 
Inspecting the dynamics of each species on both the one and the two-body level we observe that during 
the evolution, the predominantly occupied wells are one-body incoherent and mainly two-body anti-correlated with each other while 
within each well a correlated behavior, for bosons, and an anti-correlated one, for fermions, occurs.        
Furthermore, it is shown that the immiscible phase gives rise to a richer response  
when compared to the miscible phase for varying quench amplitude. 
Finally it is found that for increasing height of the potential barrier the expansion dynamics of the BF mixture is suppressed while for mass imbalanced mixtures the 
heavier component is essentially unperturbed. 

This work is organized as follows. 
In Sec. \ref{theory} we introduce our setup, the employed MB wavefunction ansatz and the basic observables of interest. 
Sec. \ref{ground} presents the ground state properties of our system. 
In Secs. \ref{immiscible} and \ref{mixed} we focus on the quench induced expansion dynamics of the BF mixture within the immiscible 
and the miscible correlated phases respectively.  
We summarize our findings and present an outlook in Sec. \ref{conclusions}. 
Appendix A presents the correlation dynamics during the expansion of the BF mixture within the 
immiscible phase and in Appendix B we show the impact of several system parameters on the expansion dynamics. 
Appendix C contains a discussion regarding the convergence of our numerical ML-MCTDHX simulations.

\section{Theoretical Framework}\label{theory} 

\subsection{Setup and Many-Body Ansatz}  

We consider a BF mixture consisting of $N_F$ spin polarized fermions and $N_B$ bosons each of mass $M$. 
This system can be to a good approximation realized by considering e.g. a mixture of isotopes of $^{7}$Li and $^{6}$Li \cite{Kempen} 
or $^{171}$Yb and $^{172}$Yb \cite{Honda,Takasu}.       
The mixture is confined in an one-dimensional optical lattice with an imposed harmonic confinement of frequency $\omega$  
and the MB Hamiltonian reads 
\begin{equation}
\begin{split}
&H=\sum_{i=1}^{N_F+N_B}[ -\frac{\hbar^2}{2M}\frac{\partial^2}{\partial x_i^2}+\frac{M}{2}\omega^2x_i^2+V_{0}sin^2(kx_i)] 
\\&+g_{FB}\sum_{i=1}^{N_F}\sum_{j=1}^{N_B}\delta(x_i^F-x_j^B)+g_{BB}\sum_{1\geq i\geq j\geq N_B}\delta(x_i^B-x_j^B).
\end{split}
\end{equation}
The lattice potential is characterized by its depth, $V_{0}$, and periodicity, $l=\pi$ (with $k = \pi/l$).   
Within the ultracold $s$-wave scattering limit the inter- and intraspecies interactions are adequately modeled by contact interactions 
scaling with the effective one-dimensional coupling strength $g_{\sigma\sigma'}$ where $\sigma,\sigma'=B, F$ for bosons or fermions respectively. 
The effective one-dimensional coupling strength \cite{Olshanii} 
${g^{1D}_{\sigma \sigma'}} =\frac{{2{\hbar ^2}{a^s_{\sigma \sigma'}}}}{{ma_ \bot ^2}}{\left( {1 - {\left|{\zeta (1/2)} \right|{a^s_{\sigma\sigma'}}}/{{\sqrt 2 {a_ \bot }}}} \right)^{ -
1}}$, where $\zeta$ denotes the Riemann zeta function. 
The transversal length scale is ${a_ \bot } = \sqrt{\hbar /{m{\omega _ \bot }}}$, and ${{\omega _ \bot }}$ is the 
frequency of the transversal confinement, while ${a^s_{\sigma\sigma'}}$ denotes the free space $s$-wave
scattering length within or between the two species.  
$g_{\sigma\sigma'}$ is tunable by ${a^s_{\sigma\sigma'}}$ via Feshbach resonances \cite{Kohler,Chin} or 
by means of ${{\omega _ \bot }}$ \cite{Olshanii,Kim}. 
$S$-wave scattering is prohibited for spinless fermions due to their antisymmetry \cite{Pethick_book,Lewenstein_book} and thus 
they are considered to be non-interacting among each other. 
The MB Hamiltonian is rescaled in units of the recoil energy ${E_{\rm{R}}} = \frac{{{\hbar ^2k^2}}}{{2M}}$. 
Then, the corresponding length, time, frequency and interaction strength scales are given in units of
${k^{ - 1}}$, $\omega_{\rm{R}}^{-1}=\hbar E_{\rm{R}}^{ - 1}$, $\omega_{\rm{R}}$ and $2E_Rk^{-1}$ respectively. 
To limit the spatial extension of our system we impose hard-wall boundary conditions at $x_\pm=\pm\frac{19}{2}\pi$.
For convenience we also shall set $\hbar = M = k = 1$ and therefore all quantities below are 
given in dimensionless units.  

Our system is initially prepared in the ground state of the MB Hamiltonian where the harmonic trap frequency is $\omega=0.1$ and the lattice depth $V_{0}=3$. 
Due to the imposed harmonic trap initially the mixture experiences a localization tendency towards the central wells which is stronger for decreasing $g_{BB}$. 
To induce the dynamics we instantaneously change at $t=0$ the trapping frequency $\omega$ to lower values and let the system evolve in time. 
Note that reducing $\omega$ predominantly favors the tunneling of both components to the outer wells as the corresponding energy offset 
between distinct wells becomes smaller. 
In this way, after the quench the mixture is prone to expand. 

To solve the underlying MB Schr{\"o}dinger equation we employ ML-MCTDHX \cite{MLX,Cao_ML}.   
The latter, in contrast to the mean-field (MF) approximation, relies on expanding the MB wavefunction in a 
time-dependent and variationally optimized basis, enabling us to take into account inter and intraspecies correlations. 
To include interspecies correlations, we first introduce $M$ distinct species functions for each component 
namely $\Psi^{\sigma}_k (\vec x^{\sigma};t)$ where $\vec x^{\sigma}=\left( x^{\sigma}_1, \dots, x^{\sigma}_{N_{\sigma}} \right)$ 
denote the spatial $\sigma=F,B$ species coordinates and $N_{\sigma}$ is the number of $\sigma$ species atoms.     
Then, the MB wavefunction $\Psi_{MB}$ can be expressed according to the truncated Schmidt decomposition \cite{Horodecki} 
of rank $M$ 
\begin{equation}
\Psi_{MB}(\vec x^F,\vec x^B;t) = \sum_{k=1}^M \sqrt{ \lambda_k(t) }~ \Psi^F_k (\vec x^F;t) \Psi^B_k (\vec x^B;t),   
\label{Eq:WF}
\end{equation}
where the Schmidt coefficients $\lambda_k(t)$ are referred to as the natural species populations of the $k$-th species function. 
The system is entangled \cite{Roncaglia} or interspecies correlated when at least two distinct $\lambda_k(t)$ are nonzero and therefore  
the MB state cannot be expressed as a direct product of two states.   
In this entangled case, a particular fermionic configuration $\Psi_k^{F}(\vec x^F;t)$ is accompanied by a 
particular bosonic configuration $\Psi_k^{B}(\vec x^B;t)$ and vice versa. 
As a consequence, measuring one of the species states, e.g. $\Psi_{k'}^{F}$, collapses the wavefunction of the other species 
to $\Psi_{k'}^{B}$ thus manifesting the bipartite entanglement \cite{Peres,Lewenstein2}.  

Moreover in order to account for interparticle correlations each of 
the species functions $\Psi^{\sigma}_k (\vec x^{\sigma};t)$ is expanded using the determinants or permanents of $m^{\sigma}$ distinct 
time-dependent fermionic or bosonic single-particle functions (SPFs), $\varphi_1,\dots,\varphi_{m^{\sigma}}$, respectively 
\begin{equation}
\begin{split}
&\Psi_k^{\sigma}(\vec x^{\sigma};t) = \sum_{\substack{n_1,\dots,n_{m^{\sigma}} \\ \sum n_i=N}} c_{k,(n_1,
\dots,n_{m^{\sigma}})}(t)\times \\ &\sum_{i=1}^{N_{\sigma}!} {\rm sign}(\mathcal{P}_i) ^{\zeta} \mathcal{P}_i
 \left[ \prod_{j=1}^{n_1} \varphi_1(x_j;t) \cdots \prod_{j=1}^{n_{m^{\sigma}}} \varphi_{m^{\sigma}}(x_j;t) \right].  
 \label{Eq:4}
 \end{split}
\end{equation} 
Here $\zeta=0,1$ for the case of bosons and fermions respectively and $\rm{sign}(\mathcal{P}_i)$ denotes the sign of the corresponding 
permutation. 
$\mathcal{P}$ is the permutation operator exchanging the particle configuration within the SPFs. 
$c_{k,(n_1,\dots,n_{m^{\sigma}})}(t)$ are the time-dependent
expansion coefficients of a particular determinant for fermions or permanent for bosons, 
and $n_i(t)$ denotes the occupation number of the SPF $\varphi_i(\vec{x};t)$. 
Note that the bosonic subsystem is termed intraspecies correlated if more than one 
eigenvalue is substantially occupied, otherwise is said to be fully coherent \cite{mueller,penrose}. 
In the same manner, the fermionic species possesses beyond Hartree-Fock intraspecies correlations if more than $N_F$  
eigenvalues occur. 
Employing the Dirac-Frenkel variational principle \cite{Frenkel,Dirac} for the MB ansatz [see Eqs.~(\ref{Eq:WF}), (\ref{Eq:4})] 
yields the ML-MCTDHX equations of motion \cite{MLX}. 
These consist of $M^2$ linear differential equations of motion for the coefficients $\lambda_i(t)$, which are coupled to a set of 
$M[$ ${N_B+m^B-1}\choose{m^B-1}$+${m^F}\choose{N_F}$] non-linear integro-differential equations for the species functions and $m^F+m^B$ 
integro-differential equations for the SPFs.   
Finally, it is also worth mentioning that ML-MCTDHX can operate in different approximation 
orders, e.g. it reduces to the MF Gross-Pitaevskii equation in the case of $M=m^F=m^B=1$.

\subsection{Observables of Interest}

Let us next briefly introduce the main observables that will be used for the interpretation of the 
expansion dynamics on both the one- and two-body level. 
To measure the collective expansion and contraction dynamics \cite{MLY,scattering} of the $\sigma$ species atomic 
cloud we rely on the position variance 
\begin{equation}
\begin{split}
\Sigma ^2_{x,\sigma}(t)&=\bra{\Psi_{MB}(t)}\hat{x}_{\sigma}^2\ket{\Psi_{MB}(t)}- \bra{\Psi_{MB}(t)}\hat{x}_{\sigma}\ket{\Psi_{MB}(t)}^2. \label{pos_var}
\end{split}
\end{equation}
Here, $\hat{x}_{\sigma}=\int_{D}dx~x_{\sigma}~\hat{\Psi}_{\sigma}^\dagger(x)\hat{\Psi}_{\sigma}(x)$, and 
$\hat{x}^{2}_{\sigma}=\int_{D}dx~x^2_{\sigma}~\hat{\Psi}_{\sigma}^\dagger(x)\hat{\Psi}_{\sigma}(x)$ are one-body operators, with   
$\hat{\Psi}_{\sigma}(x)$ denoting the $\sigma$ species field operator, and $D$ is the spatial extent of the lattice. 
We remark that the aforementioned position variance, evaluated over the entire lattice, essentially quantifies 
a ’global breathing’ mode composed of interwell tunneling and intrawell breathing modes   
offering this way a measure for the system's dynamical response.    

To elaborate on the intensity of the resulting dynamical response for the $\sigma$ species 
we define the time-averaged position variance 
\begin{equation}
\bar{\Sigma} ^2_{x,\sigma}=\frac{1}{T}\int_{0}^{T}[\Sigma ^2_{x,\sigma}(t)-\Sigma ^2_{x,\sigma}(0)],\label{average}
\end{equation}
which describes the mean deviation of the system from it's initial (ground) state. 
$\Sigma ^2_{x,\sigma}(0)$ refers to the position variance of the $\sigma$-species for the initial state at $t=0$, 
while $T$ is the considered finite evolution time in which $\bar{\Sigma} ^2_{x,\sigma}$ has converged 
to a certain value. 

The one-body reduced density matrix of the $\sigma$ species $\rho^{(1),\sigma}(x,x';t)=\bra{\Psi_{MB}(t)}\Psi^{\dagger}_{\sigma}(x')\Psi_{\sigma}(x)\ket{\Psi_{MB}(t)}$ 
provides the probability to find a $\sigma$ species particle simultaneously at positions $x$ and $x'$ at a certain time instant $t$ 
while $\rho^{(1),\sigma}(x;t)\equiv \rho^{(1),\sigma}(x,x'=x;t)$ is the $\sigma$-species single-particle density \cite{density_matrix}. 
The eigenfunctions of the $\sigma$-species one-body density matrix, $\rho^{(1),{\sigma}}(x,x')$, are the 
so-called $\sigma$-species natural orbitals, $\phi^{\sigma}_i(x;t)$, which are normalized to their corresponding eigenvalues  
\begin{equation}
n^{\sigma}_i(t)= \int d x~ \left| \phi^{\sigma}_i(x;t) \right|^2. \label{Eq:populations}
\end{equation}
$n^{\sigma}_i(t)$ are known as the natural populations of the $\sigma$ species \cite{mueller,penrose}. 
Finally, the diagonal two-body reduced density matrix $\rho^{(2),\sigma \sigma'}(x,x';t)=\bra{\Psi_{MB}(t)}\Psi^{\dagger}_{\sigma'}
(x')\Psi^{\dagger}_{\sigma}(x)\Psi_{\sigma}(x)\Psi_{\sigma'}(x')\ket{\Psi_{MB}(t)}$ refers to the probability of 
finding two atoms located at positions $x$, $x'$ at time $t$.

 \begin{figure}[ht]
 	\centering
  	\includegraphics[width=0.45\textwidth]{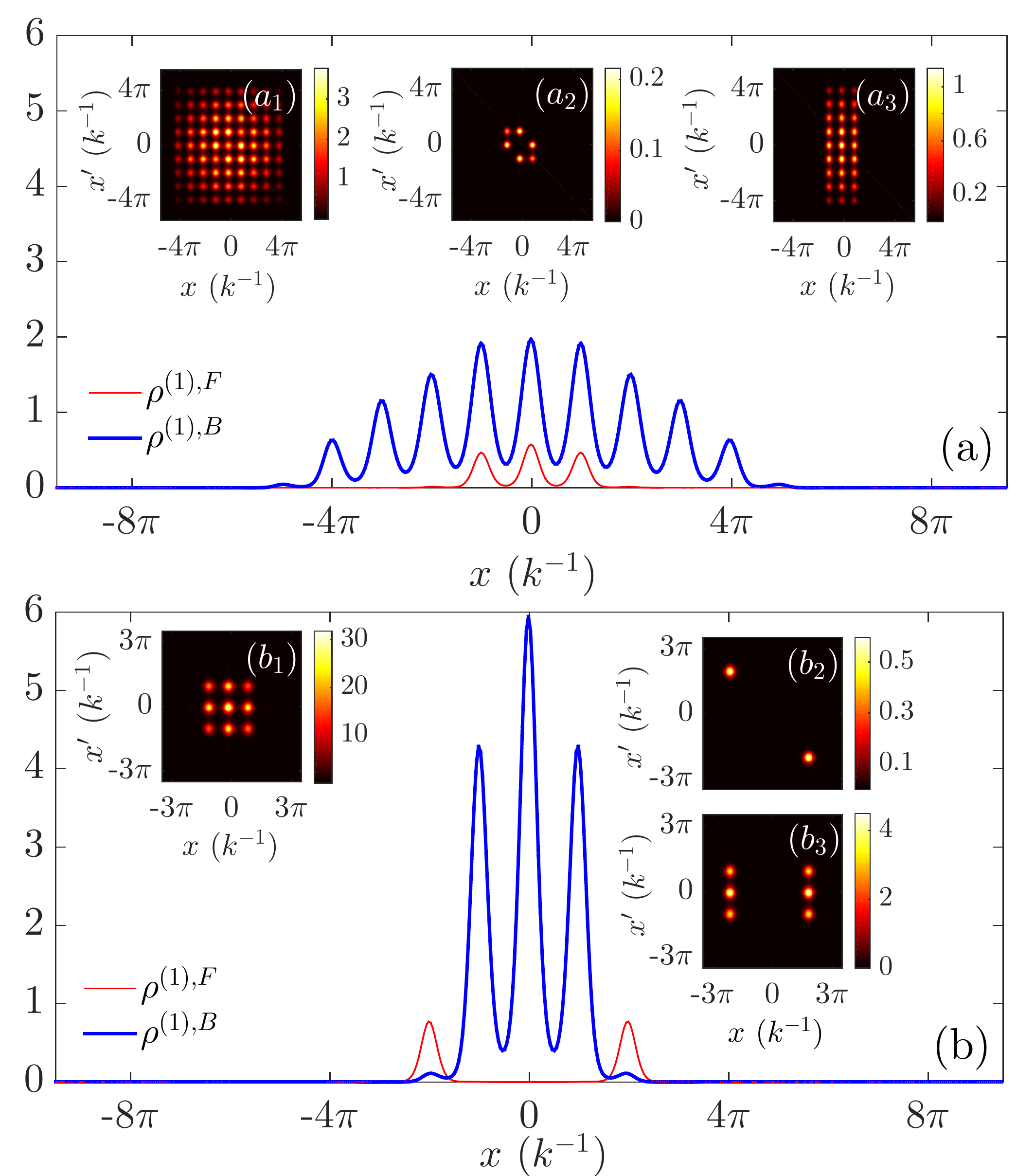}
     \caption{Fermionic (red line) and bosonic (blue line) ground state one-body densities for ($a$) $g_{BB}=1.0$, $g_{FB}=0.05$ (miscible phase) 
     and ($b$) $g_{BB}=0.05$, $g_{FB}=0.2$ (immiscible phase).   
     Insets ($a_1$), ($a_2$) show the two-body reduced density matrix of the bosons and fermions respectively for the miscible phase.   
     Insets ($b_1$), ($b_2$) show the same quantities as ($a_1$), ($b_2$) but for the immiscible phase.   
     The insets ($a_3$) and ($b_3$) depict the interspecies two-particle reduced density matrix in the miscible and immiscible regime respectively.} 
 	\label{fig:ground}
 \end{figure}

\section{Initial State Characterization}\label{ground}

Depending on the ratio between the inter ($g_{FB}$) and intraspecies ($g_{BB}$) interaction strength, the BF mixture forms two phases 
characterized by the miscibility of the bosonic and fermionic clouds \cite{Akdeniz,Akdeniz1,Das,Mistakidis_cor}.   
Here, we typically restrict ourselves to weak inter and intraspecies interactions and consider a BF mixture consisting of $N_B=20$ bosons and $N_F=2$ spin 
polarized fermions confined in a nineteen-well optical lattice. 
Tuning $\frac{g_{FB}}{g_{BB}}$ we identify different ground state configurations, namely the miscible 
and the immiscible correlated phases (see below). 
We remark that by operating within the aforementioned weak interaction regime and besides realizing the above phases, we showcase that the inclusion of correlations
is of substantial importance in order to accurately describe the expansion dynamics of the BF mixture. 
Effects of stronger interaction strengths, such as the Tonks-Girardeau regime, might be of great importance but lie beyond our scope. 

For $g_{BB}>g_{FB}$ and for $g_{BB}=1.0$ and $g_{FB}=0.05$ we realize the miscible phase where the single-particle densities  
of bosons and fermions are overlapping, see Fig. \ref{fig:ground} ($a$). 
In particular, the bosonic and fermionic single-particle densities in the three central wells overlap completely, 
while the outer wells are mainly populated by bosons. 
The broadening of the bosonic one-body density distribution is anticipated due to the strong $g_{BB}$. 
The aforementioned miscibility character of $\rho^{(1),\sigma}(x)$, favoring certain spatial regions, leads to 
the characterization of the phase as miscible. 
On the two-body level the corresponding $\rho^{(2),BB}(x,x')$ [see inset ($a_1$)] demonstrates that two bosons are likely to populate most of the available wells, 
while two fermions, see $\rho^{(2),FF}(x,x')$ in the inset ($a_2$), cannot reside in the same well but are rather delocalized over the three central wells. 
Finally, the elongated shape of $\rho^{(2),FB}(x,x')$ [see inset ($a_3$)] indicates further the miscibility of the two components within the 
three central wells and their vanishing overlap in the outer lattice wells. 

Turning to the regime of $g_{FB}>g_{BB}$, namely for $g_{BB}=0.05$ and $g_{FB}=0.2$, we enter the immiscible phase characterized by almost perfectly separated  
fermionic and bosonic single-particle densities, see Fig. \ref{fig:ground} ($b$). 
As shown, $\rho^{(1),B}(x)\neq 0$ for the three central wells (i.e. $x\in [-3\pi/2,3\pi/2]$) and 
therefore one boson is delocalized in this region.  
However, $\rho^{(1),F}(x) \neq 0$ only for the nearest neighbors of the three central wells, namely $x\in[3\pi/2,5\pi/2]$ and $x\in[-5\pi/2,-3\pi/2]$. 
The latter indicates that each fermion is localized in one of these neighboring wells. 
The above observations are also supported by the intraspecies two-body reduced density matrices \cite{Dehkharghani}. 
Indeed $\rho^{(2),BB}(x,x')\neq0$ [see inset ($b_1$)] for the three central wells implying that it is likely for two bosons to reside 
within this spatial region.  
However, $\rho^{(2),FF}(x,x')\neq0$ [see inset ($b_2$)] only for the anti-diagonal elements that refer to the nearest neighbors 
($-5\pi/2<x<-3\pi/2$ and $3\pi/2<x<5\pi/2$) of the three central wells. 
Therefore each fermion populates only one of these wells. 
The diagonals of $\rho^{(2),FB}(x,x)$ depicted in the inset ($b_3$) are almost zero, reflecting in this way the phase separated character of the state.

\section{Quench Dynamics in the Immiscible Phase}\label{immiscible} 
 
\begin{figure}[ht]
 	\centering
  	\includegraphics[width=0.48\textwidth]{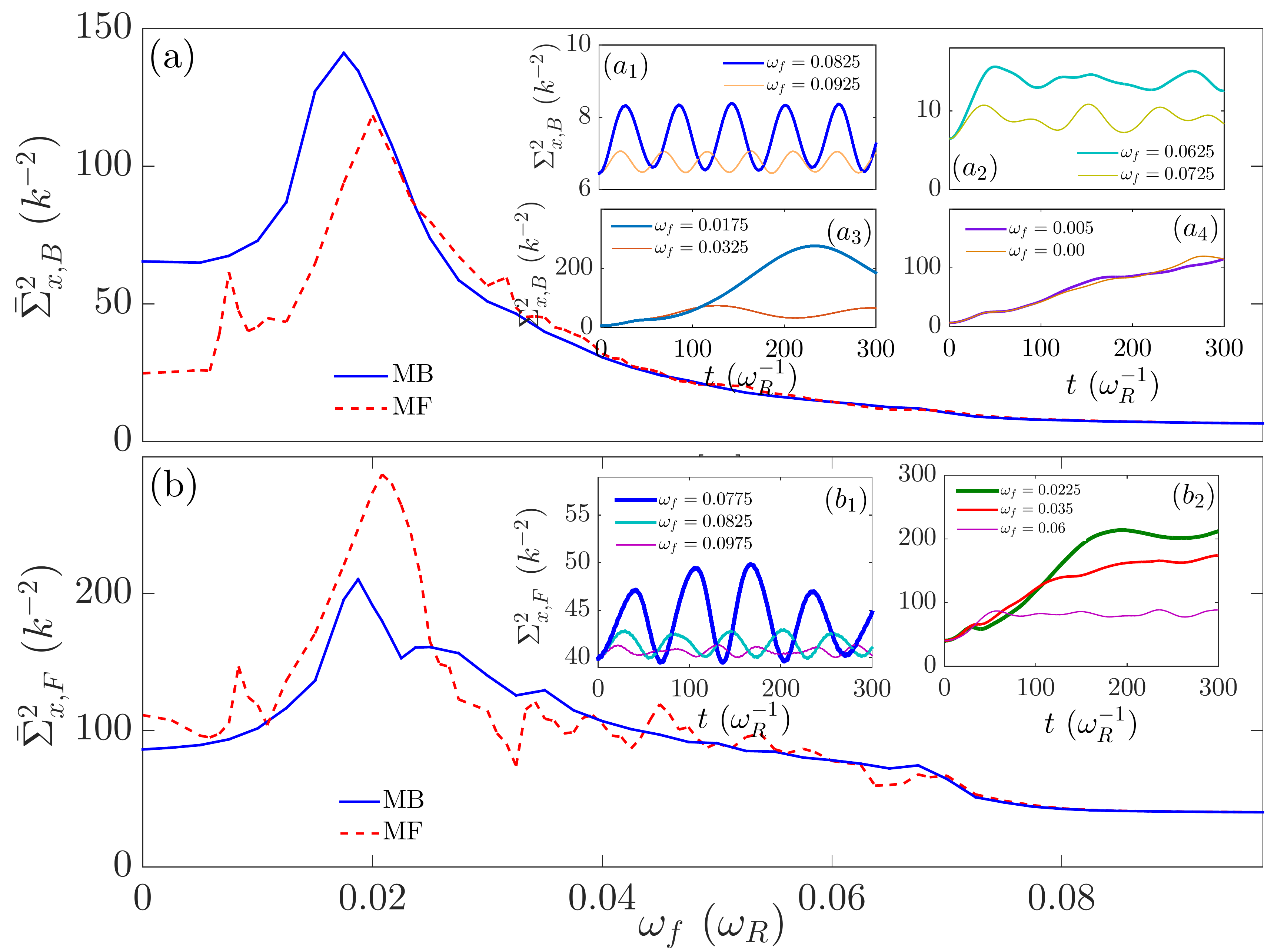}
     \caption{($a$) Bosonic and ($b$) fermionic mean variance $\bar{\Sigma} ^2_{x,\sigma}$ in the immiscible phase for varying postquench 
     harmonic trap frequency $\omega_f$. 
     ($a_1$)-($a_4$) Position variance $\Sigma_{x,B}(t)$ as a function of time within the characteristic four different bosonic response regimes. 
     ($b_1$), ($b_2$) $\Sigma_{x,F}(t)$ within the characteristic two distinct fermionic response regimes. 
     Initially the system is in the ground state of $N_B=20$ bosons and $N_F=2$ fermions with $g_{BB}=0.05$, $g_{FB}=0.2$ which are confined in a nineteen-well 
     lattice potential with an imposed harmonic trap of frequency $\omega=0.1$.}  
  	\label{fig:dynamics}
 \end{figure}

Focussing on the immiscible phase we study the expansion dynamics induced by a quench of the harmonic oscillator frequency to smaller values.  
To gain an overview of the system's mean dynamical response we resort to the $\sigma$-species time-averaged position variance 
$\bar{\Sigma} ^2_{x,\sigma}$ [see also Eq. (\ref{average})] which essentially measures the expansion strength of the atomic cloud.     
Figs. \ref{fig:dynamics} ($a$), ($b$) present $\bar{\Sigma} ^2_{x,B}$ and $\bar{\Sigma} ^2_{x,F}$ respectively with varying 
final trap frequency $\omega_f$. 
It is observed that the expansion strength strongly depends on $\omega_f$ and exhibits a maximum value in the vicinity of $\omega_f=0.0175$. 
Therefore both the bosonic and the fermionic cloud do not show their strongest expansion when completely releasing the 
harmonic trap, i.e. at $\omega_f=0$, but rather at moderate quench amplitudes. 
For either $\omega_f<0.0175$ or $\omega_f>0.0175$ an essentially monotonic decrease of $\bar{\Sigma} ^2_{x,\sigma}$ occurs 
(see also below for a more detailed description of the dynamics). 
Alterations of the overall dynamical response can be achieved by tuning 
the height of the potential barrier or the mass ratio of the two species (see Appendix B). 
The above-mentioned resonant-like behavior is reminiscent of the expansion dynamics of single-component bosons 
trapped in a composite lattice and subjected to a quench of the imposed harmonic trap from strong to weak confinement \cite{MLY}.  
In this latter case, a resonant response of the system for intermediate quench amplitudes occurs and it 
is related to the avoided crossings in the MB eigenspectrum with varying $\omega_f$. 
The occurence of the resonant-like response of the BF mixture suggests that also in the present case such avoided crossings could be responsible 
for the appearence of the maximum at $\omega_f=0.0175$. 
However, due to the large particle numbers considered herein, a direct calculation of the corresponding MB eigenspectrum is not possible. 

\begin{figure*}[ht]
 	\centering
  	\includegraphics[width=1.0\textwidth]{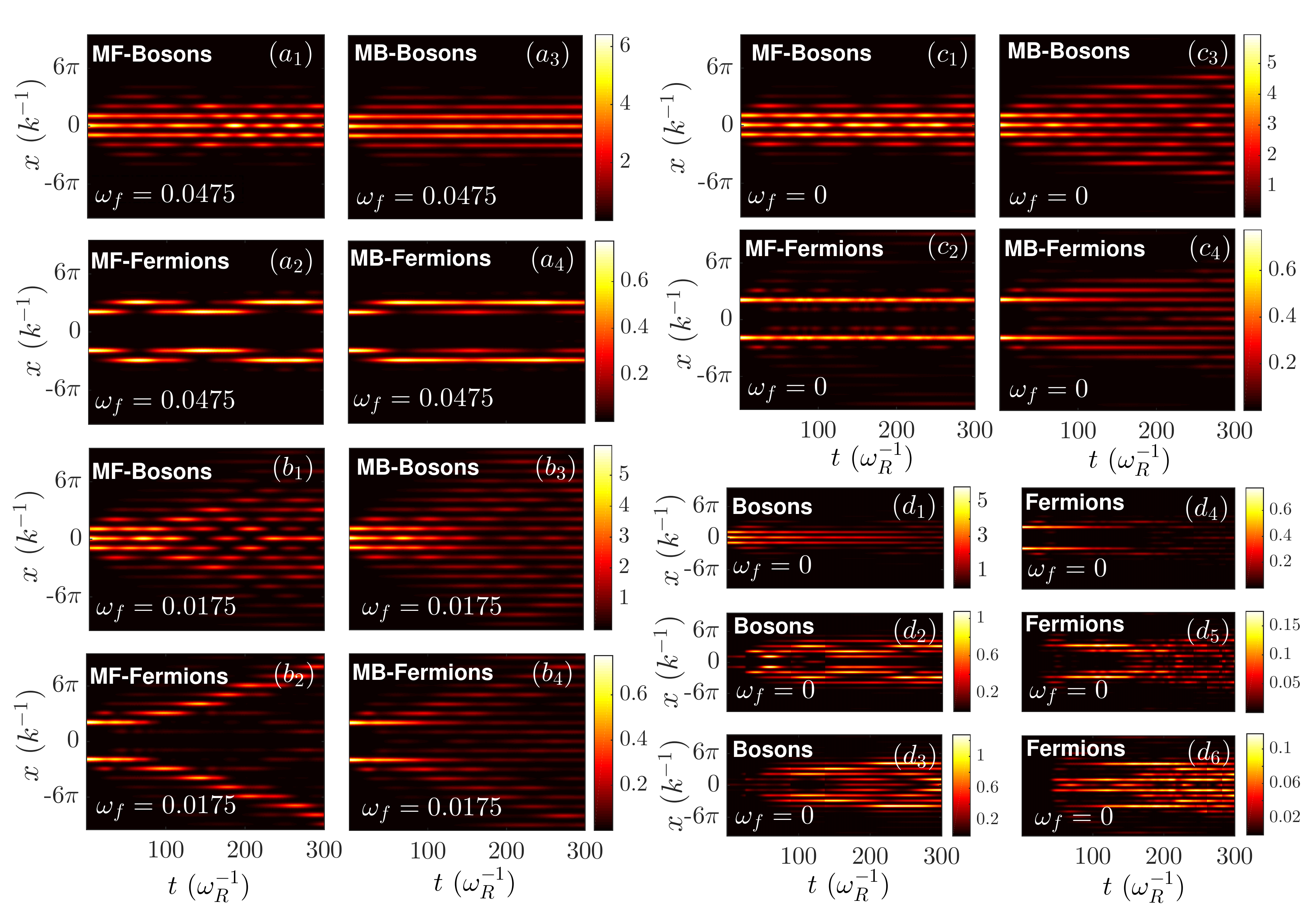}
     \caption{The one-body density evolution within the MF approach is presented in ($a_1$) and ($a_2$) for bosons and fermions respectively 
     after a quench to $\omega_f=0.0475$. 
     ($b_1$), ($b_2$) and ($d_1$), ($d_2$) present the same quantities as above but for a quench to $\omega_f=0.0175$ and $\omega_f=0.0$ respectively. 
     ($a_3$) and ($a_4$) One-body density evolution within the MB approach for bosons and fermions respectively after a quench to $\omega_f=0.0475$. 
     ($b_3$), ($b_4$) and ($c_3$), ($c_4$) present the same quantities as ($a_3$) and ($a_4$) but for a quench to $\omega_f=0.0175$ and $\omega_f=0.0$ respectively. 
     ($d_1$), ($d_2$), ($d_3$) illustrate the one-body density, in the course of the dynamics, of the first, second, re-summed third and fourth bosonic orbitals of ($c_3$).  
     ($d_4$), ($d_5$), ($d_6$) show the re-summed one-body density evolution of the first and second, third and fourth, fifth to eighth fermionic orbitals of ($c_4$). 
     The system is initialized in the ground state of $N_B=20$ bosons and $N_F=2$ fermions with $g_{BB}=0.05$, $g_{FB}=0.2$ and being confined in a nineteen-well 
    lattice potential with an imposed harmonic trap of frequency $\omega=0.1$.}  
  	\label{fig:dynamics_den}
 \end{figure*}

To elaborate in more detail on the characteristics of the dynamical response we invoke the position variance $\Sigma ^2_{x,\sigma}(t)$ [see Eq. (\ref{pos_var})] and 
the single-particle density $\rho^{(1),\sigma}(x,t)$ of the $\sigma$ species during the evolution \cite{scattering}. 
Recall that by quenching the harmonic oscillator frequency to lower values we mainly trigger the tunneling dynamics towards the outer lattice wells as their 
corresponding energy offset is reduced. 
Focussing on the bosonic species we can identify four distinct response regimes 
each one exhibiting a characteristic expansion, see Figs. \ref{fig:dynamics} ($a_1$)-($a_4$).  
Within the first regime located at $0.0775 \leq \omega_f \leq 0.1$ the bosonic cloud undergoes a regular periodic expansion 
and contraction dynamics, see the oscillatory behavior of 
$\Sigma ^2_{x,B}(t)$ in Fig. \ref{fig:dynamics} ($a_1$), being identified as a global breathing mode \cite{MLY,breath}.      
The oscillation amplitude (frequency) of $\Sigma ^2_{x,B}(t)$ increases (decreases) for smaller $\omega_f$'s lying within this region. 
In the second response regime ($0.0525\leq\omega_{f}<0.0755$) the cloud initially expands within a short evolution time ($t<50$) and 
then performs irregular oscillations possessing multiple frequencies [Fig. \ref{fig:dynamics} ($a_2$) and Fig. \ref{fig:dynamics_den} ($a_3$)]. 
The third response regime ($0.015\leq\omega_{f}\leq0.05$) is characterized by an initial expansion of the bosons until a maximum value is reached.   
Then the ensemble undergoes a contraction and follow-up expansion [Fig. \ref{fig:dynamics} ($a_3$) and Fig. \ref{fig:dynamics_den} ($b_3$)]. 
For $\omega_{f}<0.015$, defining the fourth regime, the atoms strictly expand in an approximately linear manner 
[Fig. \ref{fig:dynamics} ($a_4$) and Fig. \ref{fig:dynamics_den} ($c_4$)] reaching 
a maximum value at very long evolution times $t>600$ (not shown here). 
Their expansion velocity and amplitude are significantly reduced when compared to the third response regime 
resulting in this way in the smaller expansion strength shown in Fig. \ref{fig:dynamics} ($a$).   

Turning to the fermionic subsystem we can realize two different response regimes, see Figs. \ref{fig:dynamics} ($b_1$), ($b_2$).  
The first occurs within the same range of $\omega_f$'s as the corresponding bosonic one and $\Sigma ^2_{x,F}(t)$ performs 
regular oscillations [Fig. \ref{fig:dynamics} ($b_1$)].   
The second one appears for $\omega_f< 0.0775$ thus covering the range of quench amplitudes that lead to the second, 
third and fourth bosonic response regimes. 
Here $\Sigma ^2_{x,F}(t)$ increases monotonically for a short evolution time, reaching a maximum around 
which it oscillates with a small amplitude. 
To further visualize the dynamics of the mixture we inspect $\rho^{(1),F}(x,t)$. 
It is observed that for $\omega_f >0.03$ the bosons mainly bunch within the three central wells 
forming a material barrier \cite{Pflanzer_mat,Pflanzer_mat1} that prevents the fermions to tunnel into the inner central wells, see e.g. Fig. \ref{fig:dynamics_den} ($a_4$). 
Then the fermions perform tunneling oscillations between the two outer nearest neighboring wells located at $-9\pi/2<x<-5\pi/2$ and $5\pi/2<x<9\pi/2$.   
On the contrary for $\omega_f <0.03$ the bosons undergoe a strong expansion over the whole extent of the lattice thus allowing the fermions 
to diffuse via tunneling [Figs. \ref{fig:dynamics_den} ($b_4$) and ($c_4$)].  

\begin{figure*}[ht]
	\centering
 	\includegraphics[width=1.0\textwidth]{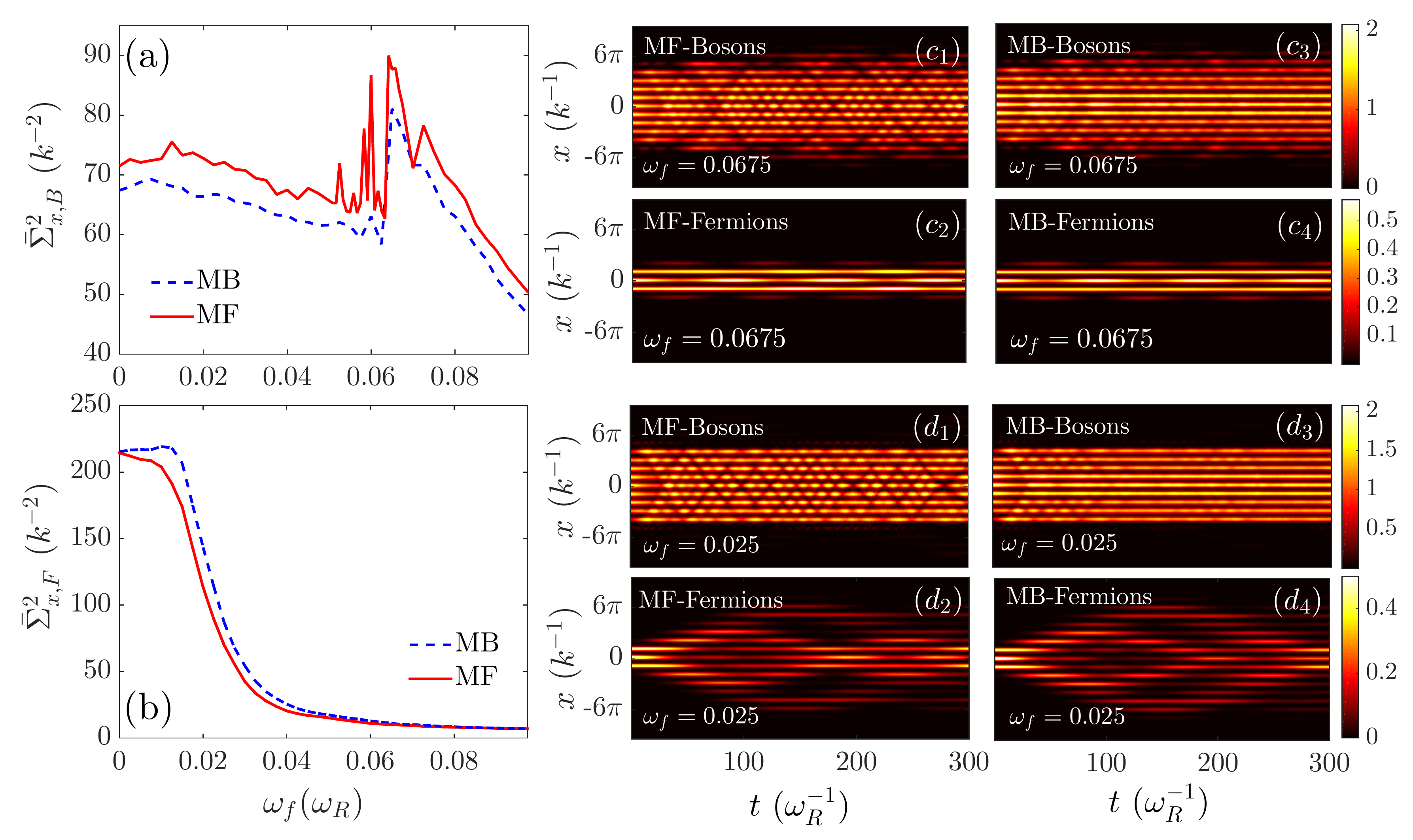}
    \caption{($a$) Bosonic and ($b$) fermionic mean variance $\bar{\Sigma} ^2_{x,\sigma}$ of a BF mixture in the miscible phase for varying postquench 
    harmonic trap frequency $\omega_f$. 
    ($c_1$), ($c_2$) Evolution of the one-body density within the MF approach for the constituting bosons and fermions respectively following a quench to $\omega_f=0.0675$. 
    ($d_1$), ($d_2$) The same as above but for $\omega_f=0.025$. 
    ($c_3$), ($c_4$) One-body density evolution within the MB approach for the bosons and fermions respectively following a quench to $\omega_f=0.0675$. 
    ($d_3$), ($d_4$) The same as above but for $\omega_f=0.025$. 
    The system is initialized in the ground state of $N_B=20$ bosons and $N_F=2$ fermions with $g_{BB}=1$, $g_{FB}=0.05$, being confined in a nineteen-well 
    lattice potential with an imposed harmonic trap of frequency $\omega=0.1$.} 
	\label{fig:dynamics_mixed}
\end{figure*}

\subsection{Identification of the Many-Body Characteristics} 

To infer about the MB nature of the above-mentioned response regimes we perform a comparison with the corresponding quench induced dynamics 
obtained within the MF (single-orbital) approximation. 
In the latter case $\bar{\Sigma} ^2_{x,B}$ for varying $\omega_f$, see Fig. \ref{fig:dynamics} ($a$), shows a qualitatively similar 
behavior to the MB case. 
However, the MF result predicts a displaced response maximum to larger values of $\omega_f$ and the existence of a secondary maximum at 
$\omega_f=0.0075$ which is suppressed in the presence of correlations.  
Comparing $\bar{\Sigma} ^2_{x,B}$ in the MB and the single-orbital approximation we can deduce that for large 
quench amplitudes ($\omega_f < 0.02$) the expansion strength is strongly suppressed in the latter case.   
Moreover the third and fourth bosonic response regimes identified within the MB approach are greatly altered in the MF realm.  
For instance, the slow monotonic expansion of the cloud in the fourth regime [see e.g. $\rho^{(1),B}(x,t)$ in Fig. \ref{fig:dynamics_den} ($c_3$)] 
is substituted by regular tunneling oscillations of the bosons in the five central wells [Fig. \ref{fig:dynamics_den} ($c_1$)]. 
Moreover MF fails to adequately capture the tunneling dynamics. 
This latter observation is clearly imprinted in the one-body density evolution presented e.g. in Figs. \ref{fig:dynamics_den} ($b_1$) and ($b_3$). 
Additionally here, significant deviations, not resolvable by inspecting $\bar{\Sigma} ^2_{x,B}$ between the two approaches, are also present, 
compare for instance Figs. \ref{fig:dynamics_den} ($a_1$) and ($a_3$). 
A careful inspection of $\rho^{(1),B}(x,t)$ reveals that in the MB scenario for $\omega_f<0.0325$ a diffusive tendency of the bosons over the entire lattice takes place 
for long evolution times, see Fig. \ref{fig:dynamics_den} ($b_3$) and ($c_3$). 

Turning to the fermionic component, and in contrast to the bosonic case, the expansion strength $\bar{\Sigma} ^2_{x,F}$ is enhanced 
in the MF approximation [Fig. \ref{fig:dynamics} ($b$)] when compared to the MB scenario for large quench amplitudes namely $\omega_f<0.025$. 
This increase of $\bar{\Sigma} ^2_{x,F}$ can be attributed to the supression of the tunneling processess 
towards the inner central wells and a dominant outward spreading, see e.g. Fig. \ref{fig:dynamics_den} ($b_2$). 
For $\omega_f<0.025$ the MB approach predicts a strong delocalization of the two fermions over the entire lattice 
for large evolution times ($t>250$) with almost all tunneling processess being damped [see e.g. Figs. \ref{fig:dynamics_den} ($b_4$) and ($c_4$)]. 
This result is in direct contrast to what it is observed in the MF case. 
Here, the fermions show an expansion being characterized by two almost localized density branches that mainly tunnel to 
the outer wells [Fig. \ref{fig:dynamics_den} ($b_2$)] while being almost localized close to the central wells at all times 
for $\omega_f=0$ [Fig. \ref{fig:dynamics_den} ($c_2$)]. 
A further discussion regarding the correlation dynamics of the BF mixture on both the one- and two-body level is provided in Appendix A. 

To gain a deeper understanding on the underlying microscopic properties of the MB dynamics we next inspect the single-particle density evolution of the participating  
orbitals $|\phi^{\sigma}_i(x,t)|^2$ after quenching to $\omega_f=0$.    
Figs. \ref{fig:dynamics_den}  ($d_1$)-($d_3$) present the corresponding single-particle densities of all four bosonic orbitals.   
The first and predominantly contributing orbital [Fig. \ref{fig:dynamics_den} ($d_1$)] shows almost no expansion and a suppressed 
tunneling dynamics within the five middle wells. 
The latter behavior resembles to a certain extent the single-particle density evolution within the MF approach, see also Fig. \ref{fig:dynamics_den} ($c_1$). 
On the other hand, the second [Fig. \ref{fig:dynamics_den} ($d_2$)], as well as the re-sumation of the third and the fourth 
orbital densities [Fig. \ref{fig:dynamics_den} ($d_3$)] indicate an expansion of the bosonic cloud over the entire lattice.  
Therefore these contributions are responsible for the above-described broader one-body density distribution of the bosons 
along the lattice in the MB (compared to MF) case. 

To analyze also the fermionic motion we next examine the single-particle densities of the eight fermionic orbitals, see Figs. \ref{fig:dynamics_den} ($d_4$)-($d_6$). 
Recall here that due to the Pauli exclusion principle each orbital can be occupied by only one fermion  
and therefore the corresponding MF approximation requires the utilization of two orbitals. 
The re-summed density of the first two fermionic orbitals [Fig. \ref{fig:dynamics_den} ($d_4$)] for $t<120$ presents the evolution of two almost 
localized single-particle density branches located at $x\to[3\pi/2,5\pi/2]$ and $x\to[-5\pi/2,-3\pi/2]$ respectively. 
Notice here the resemblance to the corresponding MF density [Fig. \ref{fig:dynamics_den} ($c_2$)] for $t<120$. 
However, for longer evolution times these density branches move towards the inner central lattice wells. 
In contrast to the above, the re-summed single-particle densities of every two consecutively occupied orbitals 
[Fig. \ref{fig:dynamics_den} ($d_5$) and ($d_6$)] exhibit a delocalization along the system. 
Therefore the diffusive behavior of the fermions during the MB expansion is mainly caused by the presence of 
these higher-lying orbitals.

\section{Quench Dynamics in the Miscible Phase}\label{mixed} 

To identify the impact of the initial phase on the expansion dynamics we next examine the response of a BF mixture, that initially resides within 
the miscible phase (with $g_{BB}=1$ and $g_{FB}=0.05$, see also Sec. \ref{ground}), following a quench of the imposed harmonic trap from strong to weak confinement $\omega_f$. 
The corresponding expansion strength of the $\sigma$-species cloud measured via $\bar{\Sigma} ^2_{x,\sigma}$ for varying $\omega_f$ 
is presented in Figs. \ref{fig:dynamics_mixed} ($a$), ($b$). 
$\bar{\Sigma} ^2_{x,B}$ increases within the interval $0.065<\omega_f<0.1$ for decreasing $\omega_f$ 
and then exhibits a decreasing behavior up to $\omega_f=0.0625$ below which it shows a slightly increasing tendency up to $\omega_f=0$. 
To visualize the emergent bosonic response we resort to the one-body density evolution $\rho^{(1),B}(x,t)$. 
The dynamical expansion of the bosonic cloud is mainly suppressed for almost every $\omega_f$ [e.g. see Fig. \ref{fig:dynamics_mixed} ($d_3$)], 
except for $0.065<\omega_f<0.072$ a region in which it becomes non-negligible [Fig. \ref{fig:dynamics_mixed} ($c_3$)].  
Instead of an expansion the bosons tunnel between the initially (at $t=0$) occupied wells and reach an almost steady state configuration 
for long evolution times [Fig. \ref{fig:dynamics_mixed} ($c_3$) and ($d_3$)].   
Despite the aforementioned triggered tunneling modes, the bosonic density reveals a maximal occupation of the 
three central wells during the dynamics [Fig. \ref{fig:dynamics_mixed} ($c_3$) and ($d_3$)]. 
To identify the effect of correlations on the bosonic expansion we compare these findings to the MF approximation. 
The mean expansion strength, $\bar{\Sigma} ^2_{x,B}$, is similar to what MB theory predicts but overall shifted to larger values [Fig. \ref{fig:dynamics_mixed} ($a$)]. 
This shift is caused by the absence of the density bunching [e.g. see Fig. \ref{fig:dynamics_mixed} ($c_1$), ($d_1$)] within the three middle wells that occurs 
in the MB scenario, leading in turn to the smaller $\bar{\Sigma} ^2_{x,B}$ observed.  
Notice also here the highly fluctuating behavior of $\bar{\Sigma} ^2_{x,B}$ around $\omega_f=0.06$ which suggests the presence of several response resonances 
that are absent in the MB case.   
Furthermore, in the MF dynamics an enhanced interwell tunneling is observed when compared to the MB case that remains robust during the evolution 
[see Figs. \ref{fig:dynamics_mixed} ($c_1$), ($c_3$) and ($d_1$), ($d_3$)].   

In contrast to bosons, a dramatic (slight) increase of the fermionic mean variance $\bar{\Sigma} ^2_{x,F}$ occurs 
for $\omega_f<0.04$ ($0.04<\omega_f<0.1$) [Fig. \ref{fig:dynamics_mixed} ($b$)]. 
This latter behavior of $\bar{\Sigma} ^2_{x,F}$ essentially designates the fermionic expansion strength for distinct $\omega_f$'s which can 
be better traced in $\rho^{(1),F}(x,t)$, see Fig. \ref{fig:dynamics_mixed} ($c_4$), ($d_4$). 
Indeed, for small quench amplitudes, i.e. $0.04<\omega_f<0.1$, the fermions expand only slightly [Fig. \ref{fig:dynamics_mixed} ($c_4$)]. 
However, for $\omega_f>0.04$ they strongly expand reaching the edges of the surrounding bosonic cloud [Fig. \ref{fig:dynamics_mixed} ($d_3$)] 
where they are partly transmitted and partly reflected moving back towards the central wells [Fig. \ref{fig:dynamics_mixed} ($d_4$)].  
The same overall phenomenology also holds for the MF case as it is evident by inspecting both 
$\bar{\Sigma} ^2_{x,F}$ [Fig. \ref{fig:dynamics_mixed} ($b$)] and $\rho^{(1),F}(x,t)$ 
[compare Figs. \ref{fig:dynamics_mixed} ($c_2$), ($c_4$) and ($d_2$), ($d_4$)].  
This similarity can be attributed to the weak interspecies interactions, $g_{FB}=0.05$, which in turn result in reduced 
interspecies correlations within this miscible regime of interactions.

\section{Conclusions}\label{conclusions} 

We have investigated the ground state properties and in particular the many-body expansion dynamics 
of a weakly interacting BF mixture confined in an one-dimensional optical lattice with a superimposed harmonic trap. 
Tuning the ratio between the inter- and intraspecies interaction strengths we have realized distinct ground state 
configurations, namely the miscible and immiscible phases. 
These phases are mainly characterized by a complete or strongly suppressed overlap of the bosonic 
and fermionic single-particle density distributions respectively. 

To induce the dynamics we perform a quench from strong to weak confinement and examine the resulting dynamical response  
within each of the above-mentioned phases for varying final harmonic trap frequencies. 
It is observed that each phase exhibits a characteristic response composed by an overall expansion of both atomic clouds and 
an interwell tunneling dynamics which can be further manipulated by adjusting the quench amplitude. 
Focussing on the immiscible phase a resonant-like response of both components occurs at moderate quench amplitudes in contrast 
to what it is expected upon completely switching off the imposed harmonic trap. 
A careful inspection of the BF mixture expansion dynamics reveals the existence of different bosonic response regimes accompanied 
by a lesser amount of fermionic ones for decreasing confinement strength. 
In particular, we find that for varying quench amplitude the bosons either perform a breathing dynamics or solely expand while the 
fermions tunnel between the nearest neighbor outer wells being located at the edges of the bosonic cloud or show a delocalized 
behavior over the entire lattice respectively. 
To identify the many-body characteristics of the expansion dynamics, we compare our findings to the mean-field approximation 
where all particle correlations are neglected. 
Here, it is shown that in the absence of correlations the tunneling dynamics of both components cannot be adequately captured, 
the bosonic expansion is suppressed and the diffusive character of the fermions is replaced by an expansion of two almost localized density 
branches to the outer wells for large quench amplitudes.  
These deviations are further elucidated by studying the evolution of the distinct orbitals used, 
with the first one resembling the mean-field approximation and the higher-orbital contributions are responsible 
for the observed correlated dynamics. 
Finally, investigating the one and two-body coherences for each species we observe that during the evolution the predominantly occupied wells 
are one-body incoherent and two-body anti-correlated among each other while within each well a correlated behavior for bosons and an anti-correlated one for fermions occurs.  

Within the miscible phase the dynamical response of the BF mixture is greatly altered. 
The bosonic expansion is significantly suppressed when compared to the immiscible phase and the 
bosons perform interwell tunneling reaching an almost steady state for long evolution times. 
The fermions, on the other hand, expand. 
When reaching the edges of the surrounding bosonic cloud they are partly transmitted and partly reflected back towards the central wells. 
Neglecting correlations the bosonic tunneling dynamics is found to be enhanced and remains undamped, 
during the evolution in contrast to the many-body approach, while the fermionic expansion resembles adequately the many-body case. 

As a final attempt we have examined the dependence of the BF mixture expansion strength on the potential barrier height 
and the mass imbalance between the two components. 
We find that upon increasing the height of the potential barrier the expansion dynamics is suppressed 
while for mass imbalanced mixtures the heavy (bosonic) component remains essentially unperturbed. 

There are several interesting directions that one might pursue in future studies. 
A straightforward one would be to explore the dynamics of the BF mixture setup but now induced 
by a quench from strong to weak confinement only for the fermionic ensemble thus letting the bosons unaffected.  
In this setting, the bosonic system may act as a filter which absorbs completely or partly the momentum of the expanded 
fermions depending on the quench amplitude.  
Yet another intriguing prospect is to examine the dynamics of a dipolar BF mixture under the quench protocol considered herein, and 
investigate the distinct response regimes that appear for varying quench amplitude or  
initial phase so as to explore the possibility to induce a ballistic expansion.

\appendix

\section*{Appendix A: Correlation Dynamics in the Immiscible Phase} 

To further elaborate on the MB nature of the expansion dynamics of the BF mixture within the immiscible phase we study the emergent 
correlation properties of the system on both the one- and two-body level. 
To estimate the degree of spatial first order coherence during the expansion dynamics, we employ \cite{Naraschewski}  
\begin{equation}
g^{(1),\sigma}(x,x';t)=\frac{\rho^{(1),\sigma}(x,x';t)}{\sqrt{\rho^{(1),\sigma}(x;t)\rho^{(1),\sigma}(x';t)}}, \label{one_body_coherence}
\end{equation} 
where $\rho^{(1),\sigma}(x,x';t)=\bra{\Psi_{MB}(t)}\Psi^{\dagger}_{\sigma}(x')\Psi_{\sigma}(x)\ket{\Psi_{MB}(t)}$ is the one-body reduced density 
matrix of $\sigma$ species. 
$|g^{(1),\sigma}(x,x';t)|^2$ takes values within the range $[0,1]$, while 
a spatial region with $|g^{(1),\sigma}(x,x';t)|^2=0$ ($|g^{(1),\sigma}(x,x';t)|^2=1$) is referred 
to as fully incoherent (coherent). 

\begin{figure}[ht]
                \centering
		\includegraphics[width=0.5\textwidth]{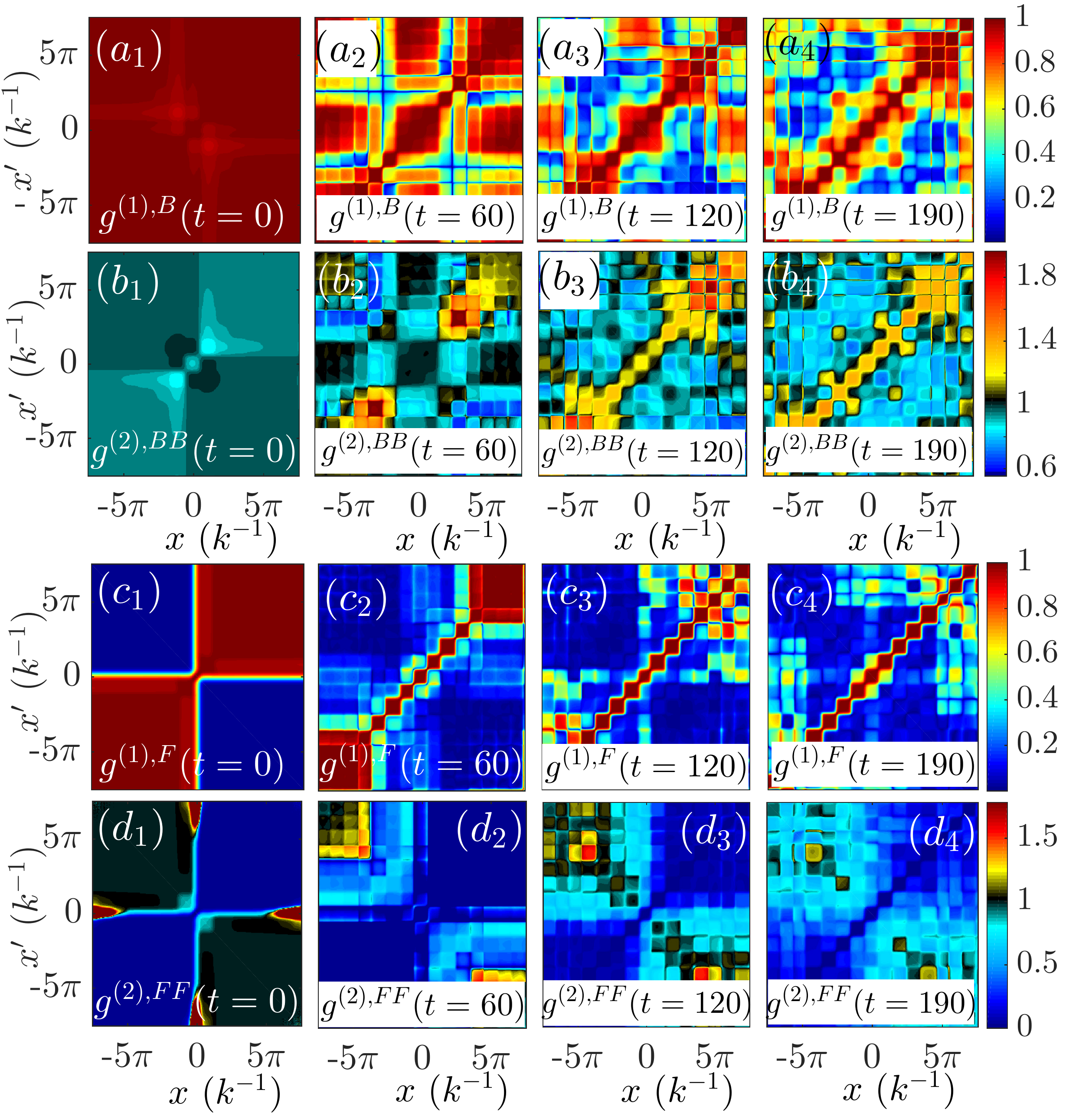}
	\caption{($a_1$)-($a_4$) One-body coherence function $g^{(1),B}(x,x';t)$ shown for different time 
  	instants (see legends) during the expansion dynamics within the immiscible phase ($g_{BB}=0.05$, $g_{FB}=0.2$). 
  	($c_1$)-($c_4$) The same as above but for $g^{(1),F}(x,x';t)$. 
  	($b_1$)-($b_4$) Snapshots of the corresponding two-body bosonic coherence function $g^{(2),BB}(x_1,x_2';t)$. 
  	($d_1$)-($d_4$) The same as before but for $g^{(2),FF}(x_1,x_2';t)$ of the fermionic component. 
  	The BF mixture consists of $N_B=20$ bosons and $N_F=2$ fermions confined in a nineteen-well optical lattice with an 
	imposed harmonic trap with initial frequency $\omega=0.1$.}	
	\label{fig:coherence}
\end{figure}

Figs. \ref{fig:coherence} ($a_1$)-($a_4$) and ($c_1$)-($c_4$) present $g^{(1),B}(x,x';t)$ and $g^{(1),F}(x,x';t)$ respectively for distinct 
time instants during evolution after quenching the system to $\omega_f=0$. 
Referring to the bosonic component we observe that at $t=0$ (ground state) the ensemble is almost  
perfectly one-body coherent as $g^{(1),B}(x,x';t)\approx1$ everywhere [Fig. \ref{fig:coherence} ($a_1$)]. 
However upon quenching this situation changes drastically and a substantial loss of coherence 
in the off-diagonal elements of $g^{(1),B}(x,x';t)$ occurs throughout the dynamics, see Figs. \ref{fig:coherence} ($a_2$)-($a_4$).  
The latter implies that the quench operation and loss of coherence go hand in hand. 
In particular, we can identify three different spatial regions [see for instance Fig. \ref{fig:coherence} ($a_3$)] 
in which the coherence is mainly preserved. 
The first one contains the three central wells ($x,x'\in[-3\pi/2,3\pi/2]$) while the other two regions, not fixed throughout the dynamics, lie in the outer wells 
(e.g. at $t=120$ they are located at $x,x'\in[3\pi/2,9\pi/2]$ and $x,x'\in[-3\pi/2,-9\pi/2]$ respectively).   
Furthermore the aforementioned regions coincide with the areas where the different orbital densities   
contribute significantly to the MB density [Fig. \ref{fig:dynamics_den} ($d_1$)-($d_3$)].  
Indeed, as time evolves the first region exhibits a contraction [Fig. \ref{fig:coherence} ($a_3$)] and an  
expansion [Fig. \ref{fig:coherence} ($a_4$)], resembling the tunneling oscillations 
in the first orbital [Fig. \ref{fig:dynamics_den} ($d_1$)].    
The second and third regions travel towards the outer wells [Fig. \ref{fig:coherence} ($a_4$)] 
in the course of the dynamics, reflecting the expansion of the second, third and fourth 
orbital densities [Fig. \ref{fig:dynamics_den} ($d_2$), ($d_3$)]. 
Finally, a significant loss of coherence takes place ($g^{(1),B}(x,x';t)\approx0.2$) between 
each two of the above-mentioned regions.   

To infer about the degree of spatial second order coherence we study the normalized two-body 
correlation function \cite{density_matrix} 
\begin{equation}
\begin{split}
g^{(2),\sigma \sigma'}(x,x';t)=\frac{\rho^{(2),\sigma \sigma'}(x,x';t)}{\rho^{(1),\sigma}(x;t)\rho^{(1),\sigma'} 
(x';t)},\label{two_body_coherence}
\end{split}
\end{equation} 
where $\rho^{(2),\sigma \sigma'}(x,x';t)=\bra{\Psi_{MB}(t)}\Psi^{\dagger}_{\sigma'}
(x')\Psi^{\dagger}_{\sigma}(x)\Psi_{\sigma}(x)\\\Psi_{\sigma'}(x')\ket{\Psi_{MB}(t)}$ 
is the diagonal two-body reduced density matrix. 
When referring to the same (different) species, i.e. $\sigma=\sigma'$ ($\sigma \not=\sigma'$), 
$g^{(2),\sigma \sigma'}(x,x';t)$ accounts for the intraspecies (interspecies) two-body correlations. 
A perfectly condensed MB state corresponds to $g^{(2),\sigma \sigma'}(x,x';t)=1$ and it is termed fully 
second order coherent or uncorrelated. 
However, if $g^{(2),\sigma \sigma'}(x,x';t)$ takes values larger (smaller) than unity the state is said to be  
correlated (anti-correlated) \cite{granulation,density_matrix}.    

In Figs. \ref{fig:coherence} ($b_1$)-($b_4$) and ($d_1$)-($d_4$) we show $g^{(2),BB}(x,x';t)$ and $g^{(2),FF}(x,x';t)$ for 
different evolution times when quenching the system to $\omega_f=0$. 
The bosonic subsystem is initially ($t=0$) mainly characterized by weak two-body anticorrelations, 
i.e. $g^{(2),BB}(x,x';t)<1$ [Fig. \ref{fig:coherence} ($b_1$)].       
The quench gives rise to new correlation structures, see Fig. \ref{fig:coherence} ($b_2$)-($b_4$).  
For instance, a bunching tendency occurs in the diagonal elements, i.e. $g^{(2),BB}(x,x';t)>1$, indicating that it 
is probable for two bosons to reside within the same well during the dynamics. 
Most importantly we observe that each of the above-described second and third regions 
of almost perfect one-body coherence (e.g. see $x,x'\in[3\pi/2,9\pi/2]$ and $x,x'\in[-3\pi/2,-9\pi/2]$ respectively at $t=120$) 
are two-body correlated while they are mainly  
anticorrelated between each other [e.g. see Fig. \ref{fig:coherence} ($b_3$)].   
Overall the off-diagonal elements of the $g^{(2),BB}(x,x';t)$ tend to values smaller than unity, 
indicating long-range anti-correlations in the system.  
Comparing $g^{(1),B}(x,x';t)$ and $g^{(2),BB}(x,x';t)$ we can infer that when $g^{(2),BB}(x,x';t)>1$ ($g^{(2),BB}(x,x';t)<1$) 
the corresponding $g^{(1),B}(x,x';t)\approx1$ ($g^{(1),B}(x,x';t)\leq0.5$).  

In contrast to the bosons, initially ($t=0$) each fermion is localized either in the left ($-20<x<0$) or in the right ($0<x<20$) part of the lattice [see also Fig. \ref{fig:ground} ($c$)].  
Indeed, $g^{(1),F}(x,x';t)\approx1$ and $g^{(2),FF}(x,x';t=0)<1$ ($g^{(1),F}(x,x';t=0)=0$ and $g^{(2),FF}(x,x';t=0)\approx1$) within (between) the 
left and right part, see Fig. \ref{fig:coherence} ($c_1$) and ($d_1$) respectively. 
For later times ($t>0$) a significant loss of one-body coherence takes place manifested by the almost zero off-diagonal elements in 
$g^{(1),F}(x,x';t)\approx0$ throughout the evolution [Figs. \ref{fig:coherence} ($c_2$)-($c_4$)]. 
On the two-body level we observe the rise of long-range correlations between the parity symmetric expanded parts, e.g.  
$g^{(2),FF}(x=7\pi/2,x'=-7\pi/2;t)\approx1.3$ in Fig. \ref{fig:coherence} ($d_2$), ($d_3$), which transform into 
anticorrelations for long propagation times [Fig. \ref{fig:coherence} ($d_4$)]. 
Finally an anticorrelated behavior occurs within the same (i.e. right with $x,x'\in[0,6\pi]$ or left with $x,x'\in[-6\pi,0]$ in Fig. \ref{fig:coherence}) part of the 
lattice throughout the evolution, see for instance $g^{(2),FF}(x=2\pi,x'=2\pi;t)$ in Fig. \ref{fig:coherence} ($d_2$)-($d_4$).

\section*{Appendix B: Control of the Expansion Dynamics}\label{control}

Having analyzed in detail the expansion dynamics of the BF mixture within the immiscible and miscible 
correlated phases, let us discuss how the overall dynamics can be altered by adjusting certain initial 
system parameters. 
\begin{figure}[ht]
\centering
		\includegraphics[width=0.48\textwidth]{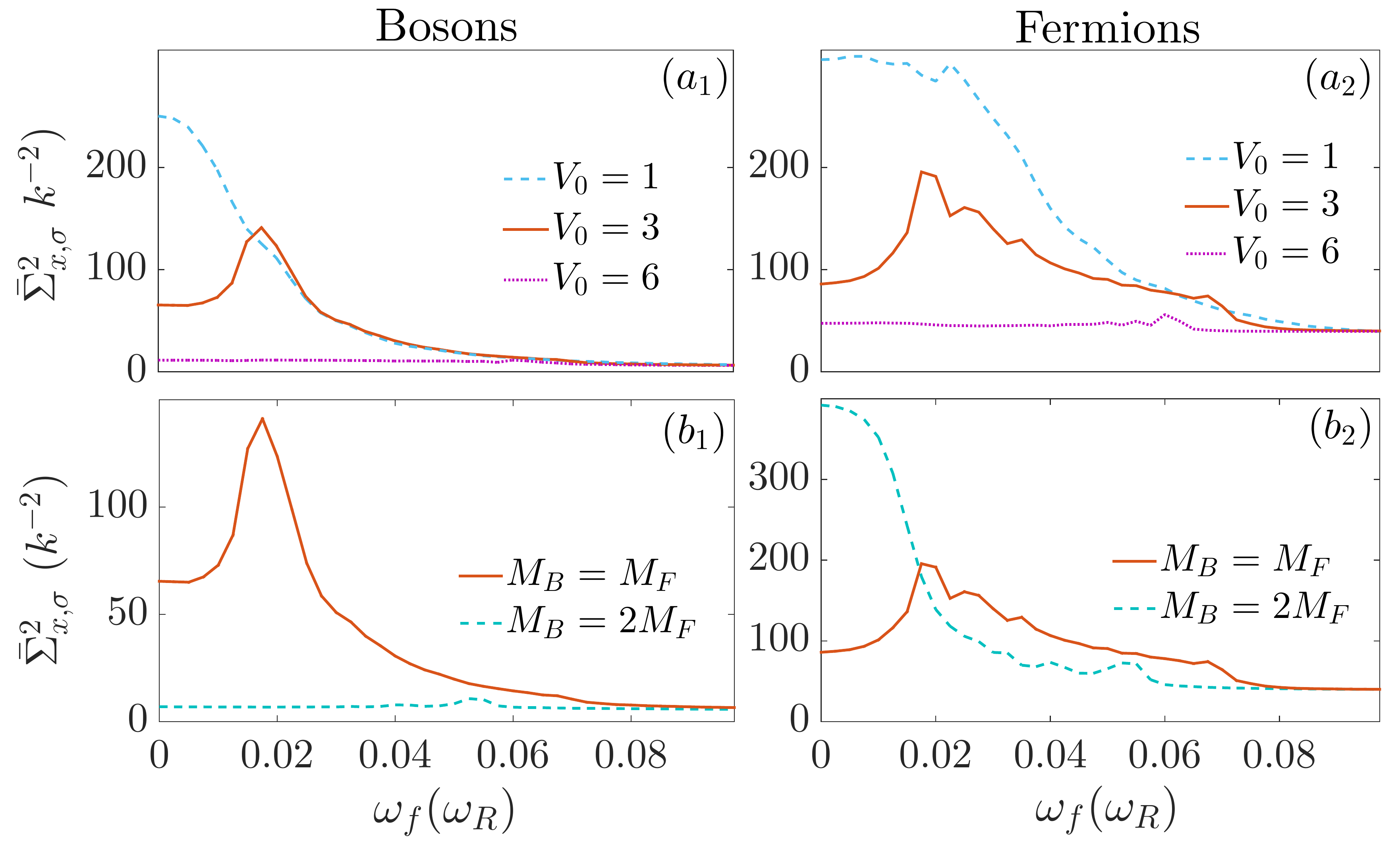}
	\caption{($a_1$), ($b_1$) Bosonic and ($a_2$), ($b_2$) fermionic mean variance $\bar{\Sigma} ^2_{x,\sigma}$ corresponding to different system parameters 
	for varying postquench frequency $\omega_f$. 
	$\bar{\Sigma} ^2_{x,\sigma}(\omega_f)$ for distinct ($a_1$), ($a_2$) potential barrier 
	heights $V_{0}$ in units of $E_R$, and ($b_1$), ($b_2$) for different mass ratios of the individual components. 
	In all cases the BF mixture consists of $N_{B}=20$ bosons, $N_F=2$ fermions and it is confined in a nineteen-well potential with an 
	imposed harmonic trap of initial frequency $\omega=0.1$. 
	The system is initialized in its ground state with $g_{BB}=0.05$ and $g_{FB}=0.2$.}	
	\label{fig:control}
\end{figure}

First we study the effect of the potential barrier height $V_{0}$ on the expansion dynamics of an ensemble that resides 
in the immiscible phase, see Fig. \ref{fig:control} ($a_1$) and ($a_2$). 
As it can be seen the corresponding expansion strength measured via $\bar{\Sigma} ^2_{x,\sigma}$ for both fermions and bosons becomes larger for 
smaller $V_{0}$ values.  
The latter is a consequence of the fact that interwell as well as overbarrier tunneling is more favorable 
for reduced barrier heights \cite{Mistakidis_cradle,Mistakidis_cradle,Jannis,Mistakidis_linear,Thies,Mistakidis_neg}.  
Note also here that the resonant expansion located at moderate quench amplitudes (see $\omega_f=0.0175$) occurs only for $V_{0}=3$.  
In contrast for $V_{0}=6$, $\bar{\Sigma} ^2_{x,\sigma}$ is almost constant for all $\omega_f$ indicating a negligible response, while 
at $V_{0}=1$, $\bar{\Sigma} ^2_{x,\sigma}$ exhibits an almost monotonic increase for decreasing $\omega_f$. 
This observation suggests that for fixed $\omega_f$ as well as inter and intraspecies interactions the expansion strength can be manipulated 
by tuning the potential barrier height. 

Another way to control the expansion dynamics is to consider a mass imbalanced BF mixture being experimentally realizable 
by using e.g. isotopes of $^{40}$K and $^{89}$Rb \cite{Chin,Wu} which possess approximately a mass ratio of 1:2. 
The system is in this case initialized in the ground state of the lattice with $g_{FB}=0.2$ and $g_{BB}=0.05$. 
Therefore it resides in the immiscible phase [see also Sec. \ref{ground}] where the two components are phase separated.  
The degree of this phase seperation increases for larger bosonic masses (results not shown here). 
Comparing a mass balanced ($M_B=M_F$) with a mass imbalanced system ($M_B=2~M_F$), we observe that the 
bosonic mass strongly influences both the fermionic and the bosonic dynamics, see Fig. \ref{fig:control} ($b_1$) and ($b_2$). 
For $M_B=2~M_F$ the bosons are essentially unperturbed for all $\omega_f$, while the fermionic expansion becomes 
significant for small $\omega_f$.  
The enhancement of $\bar{\Sigma} ^2_{x,F}$ can be explained as follows. 
First, the tunneling probability to the inner wells is surpressed due to the constantly high bosonic one-body density within 
the three central wells which essentially forms an additional material barrier \cite{Pflanzer_mat,Pflanzer_mat1}. 
Furthermore the fermionic cloud can expand ballistically, as the interspecies scattering processes in the outer wells 
are negligible since the bosonic distribution in these wells is nearly zero.  

In summary we can infer that the fermions exhibit a more pronounced expansion as compared to the bosons. 
This can be attributed to the fact that the fermions are non-interacting and as 
such they are exposed to less scattering processes when compared to bosons \cite{scattering}. 
Moreover, by tuning several of the system's parameters, allows for a control of the system's expansion dynamics in a systematic fashion.

\section*{Appendix C: Convergence of Many-Body Simulations}

In the present appendix we provide a brief overview of our numerical methodology 
and elaborate on the convergence of our results. 
ML-MCTDHX \cite{MLX} is a variational method for solving the time-dependent 
MB Schr{\"o}dinger equation of Bose-Bose \cite{Mistakidis_cor,Katsimiga_DBs}, Fermi-Fermi \cite{Cao_FF,Koutentakis_FF} and Bose-Fermi mixtures. 
The MB wavefunction is expanded with respect to a time-dependent 
variationally optimized MB basis, which enables us to capture the 
important correlation effects using a computationally feasible basis size. 
In this way, we are able to span more efficiently the relevant, for the system under consideration, subspace 
of the Hilbert space at each time instant with a reduced number of basis states when compared to expansions relying on 
a time-independent basis. 
Finally, the multi-layer ansatz for the total wavefunction allows us to account 
for intra- and interspecies correlations when simulating the dynamics of bipartite systems. 

Within our simulations, we employ a primitive basis consisting of 
a sine discrete variable representation including 475 grid points. 
The Hilbert space truncation, i.e. the order of the used approximation, 
is indicated by the considered numerical configuration space $C=(M;m^F;m^B)$. 
Here, $M=M^F=M^B$ ($m^F$, $m^B$) denote the number of species (single-particle) functions 
for each of the species. 
To maintain the accurate performance of the numerical integration for the ML-MCTDHX equations of motion 
we further ensured that $|\langle \Psi |\Psi \rangle -1| < 10^{-10}$ and
$|\langle \varphi_i |\varphi_j \rangle -\delta_{ij}| < 10^{-10}$ for the 
total wavefunction and the single-particle functions respectively. 
\begin{figure}[ht]
\centering
 	\includegraphics[width=0.48\textwidth]{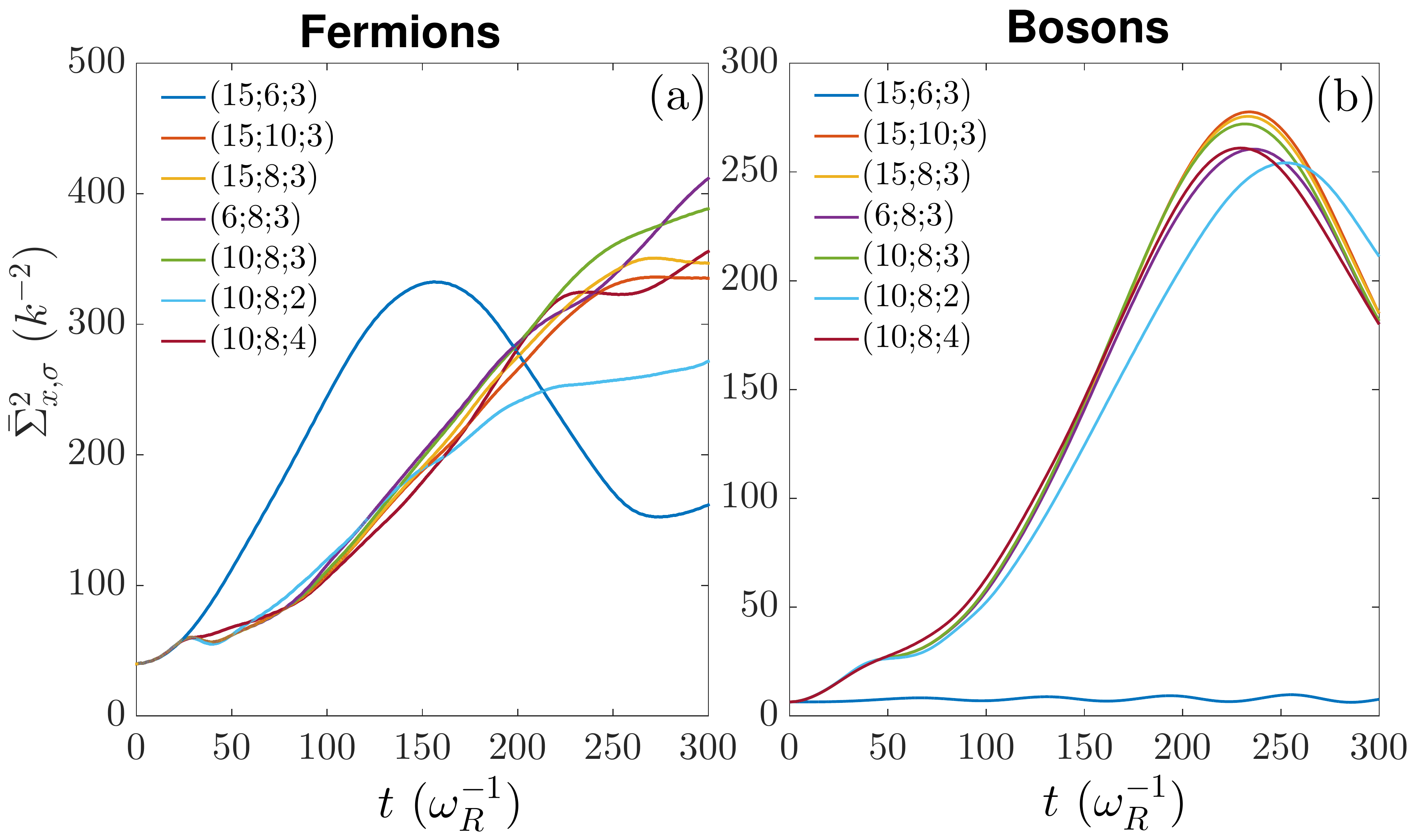}
 	\caption{ Evolution of the ($a$) fermionic and ($b$) bosonic variance $\Sigma ^2_{x,\sigma}(t)$ within 
 	the immiscible phase ($g_{FB}=0.2$ and $g_{BB}=0.05$) for different numerical configurations ($M;m^F;m^B$), see legend, 
 	following a quench to $\omega_f=0.175$. } 
 	\label{Fig:convergence}
\end{figure}

Next, let us comment on the convergence of our results upon varying the numerical configuration space $C=(M;m^F;m^B)$. 
To conclude about the reliability of our simulations, we increase the number of species functions and 
single-particle functions, thus observing a systematic convergence of our results. 
We remark that all MB calculations presented in the main text rely on the configuration $C=(10;8;4)$. 
To be more concrete in the following we demonstrate the convergence 
procedure for the position variance $\Sigma^2_{x,\sigma}(t)$ of the $\sigma$ 
species within the immiscible phase ($g_{FB}=0.2$ and $g_{BB}=0.05$) for a varying number of species or single-particle functions. 
Fig. \ref{Fig:convergence} ($a$) [($b$)] presents $\Sigma^2_{x,F}(t)$ [$\Sigma^2_{x,B}(t)$] following a quench 
of the imposed harmonic oscillator frequency from $\omega=0.1$ to $\omega_f=0.0175$. 
For reasons of completeness we remark that this quench amplitude refers to a strong response region of the system, see also Fig. \ref{fig:dynamics}.  
Regarding the number of the used species functions, $M$, we observe an adequate convergence of both the fermionic and bosonic 
variance. 
In particular, comparing the $C=(10;8;3)$ and $C=(15;8;3)$ approximations, $\Sigma^2_{x,F}(t)$ shows 
a maximal deviation of the order of $10\%$ for large propagation times $t>250$ while $\Sigma^2_{x,B}(t)$ is almost insensitive as  
the corresponding relative difference is less than $1.5\%$ throughout the evolution. 
Increasing the number of the fermionic single-particle functions, $m^F$, the maximum deviation 
observed in $\Sigma^2_{x,F}(t)$ [$\Sigma^2_{x,B}(t)$] between the $C=(15;8;3)$ and $C=(15;10;3)$ 
approximations is of the order of $4\%$ [$<1\%$]. 
Turning to the number of bosonic single-particle functions, $m^B$, the relative difference in $\Sigma^2_{x,F}(t)$ [$\Sigma^2_{x,B}(t)$] 
between the configurations $C=(10;8;3)$ and $C=(10;8;4)$ becomes at most $11\%$ [$4\%$] for large evolution times $t>230$. 
Finally, we remark that the same analysis has been performed for the convergence within the miscible regime ($g_{BB}=1.0$, $g_{FB}=0.05$) 
for increasing both the number of species, $M$, as well as the single-particle functions, $m^F$ and $m^B$ (not shown here).

\section*{Acknowledgements} 
The authors gratefully acknowledge financial support by the Deutsche Forschungsgemeinschaft 
(DFG) in the framework of the SFB 925 ``Light induced dynamics and control of correlated quantum systems''.

{}

\end{document}